\documentclass[journal]{IEEEtran}
\usepackage{graphicx}
\usepackage{hyperref}
\usepackage{adjustbox}
\usepackage{subfigure}

\usepackage{algorithmic}
\usepackage{graphicx}
\usepackage{textcomp}
\usepackage{xcolor}
\usepackage{times}
\usepackage{comment}
\usepackage{epsfig}
\usepackage{amsmath}
\usepackage{amssymb}
\usepackage{url}
\usepackage{array}
\usepackage{multirow}
\usepackage{color}
\usepackage{makecell}
\usepackage{textcomp}
\usepackage{resizegather}
\usepackage{footnote}
\usepackage{breqn}

\usepackage{fancyhdr}
\fancypagestyle{firstpage}{%
  \lfoot{\small{\textit{This paper is published in Neurocomputing Journal, Elsevier. \hspace{75mm} \copyright2020 Elsevier. 
Final paper is available at: https://doi.org/10.1016/j.neucom.2020.06.104.} 
}}
}

\begin{document}

\title{PCSGAN: Perceptual Cyclic-Synthesized Generative Adversarial Networks for Thermal and NIR to Visible Image Transformation}

\author{Kancharagunta Kishan Babu and Shiv Ram Dubey
\thanks{K.K. Babu, and S.R. Dubey are with the Computer Vision Group, \newline Indian Institute of Information Technology, Sri City, Chittoor, Andhra Pradesh - 517 646, India.
{\tt\small email: \{kishanbabu.k, srdubey\}@iiits.in}}%
}

\maketitle
\thispagestyle{firstpage}
\begin{abstract}
In many real world scenarios, it is difficult to capture the images in the visible light spectrum (VIS) due to bad lighting conditions. However, the images can be captured in such scenarios using Near-Infrared (NIR) and Thermal (THM) cameras. The NIR and THM images contain the limited details. Thus, there is a need to transform the images from THM/NIR to VIS for better understanding. However, it is non-trivial task due to the large domain discrepancies and lack of abundant datasets. Nowadays, Generative Adversarial Network (GAN) is able to transform the images from one domain to another domain. Most of the available GAN based methods use the combination of the adversarial and the pixel-wise losses (like $L_1$ or $L_2$) as the objective function for training. The quality of transformed images in case of THM/NIR to VIS transformation is still not up to the mark using such objective function. Thus, better objective functions are needed to improve the quality, fine details and realism of the transformed images. A new model for THM/NIR to VIS image transformation called Perceptual Cyclic-Synthesized Generative Adversarial Network (PCSGAN) is introduced to address these issues. The PCSGAN uses the combination of the perceptual (i.e., feature based) losses along with the pixel-wise and the adversarial losses. Both the quantitative and qualitative measures are used to judge the performance of the PCSGAN model over the WHU-IIP face and the RGB-NIR scene datasets. The proposed PCSGAN outperforms the state-of-the-art image transformation models, including Pix2pix, DualGAN, CycleGAN, PS2GAN, and PAN in terms of the SSIM, MSE, PSNR and LPIPS evaluation measures.
The code is available at \url{https://github.com/KishanKancharagunta/PCSGAN}.
\end{abstract}

\section{Introduction}
\label{introduction}
The Thermal (THM) and Near-Infrared (NIR) cameras are used to capture the images in those situations, where Visible (VIS) cameras fail to capture. The image captured in the THM/NIR domain is difficult to understand by human examiners due to the lack of information. Moreover, a large domain gap also exists between the VIS and THM/NIR images. Nowadays, the importance to match the images captured from THM/NIR to VIS is increasing due to their extensive usage in real world applications, such as for military, law enforcement, commercial, and etc. \cite{Applications}. 

The Thermal/NIR to visible image transformation has been an active research area due to great demand in real world applications. Broadly, this problem can be categorized under image generation and transformation.
Generative Adversarial Network (GAN) \cite{gan} was developed by Goodfellow et al. in 2014 to generate the images learned from the given data distribution by providing a latent vector $z$ as the input. Later on, GAN and a variation of GAN called conditional GAN (cGAN) \cite{cGAN} was proposed for image-to-image translation. It has shown a wide variety of applications, such as image manipulation \cite{AIMIM},  image super-resolution \cite{DLISR}, \cite{SISRDL}, image style transfer \cite{PLRTSTSR}, \cite{PRSISRGAN}, image-to-image transformation \cite{cGAN}, \cite{CyclicGAN}, \cite{Photo-to-Caricature}, \cite{triplefaceimage}, image in-painting \cite{multi-image-inpainting}, feature detection in images \cite{saliency} and etc. Recent developments are Cyclic Synthesized GAN (CSGAN) \cite{csgan}, Cyclic Discriminative GAN (CDGAN) \cite{cdgan}, Style-Based Generator GAN (SG-GAN) \cite{karras2019style}, and Generative adversarial minority oversampling (GAMO) \cite{mullick2019generative}.

Isola et al. have proposed Pix2pix \cite{cGAN}, a common framework for image-to-image transformation conditioned on the input image suitable only for paired dataset. Wang et al. have extended Pix2Pix to Perceptual Adversarial Network (PAN) \cite{PerceptualGAN}. They have used the perceptual loss between the features of generated and target images. Zhu et al. and Yi et al. have introduced CycleGAN \cite{CyclicGAN} and DualGAN \cite{dualGAN}, respectively, by adding a constraint between the real and the cycled images. Wang et al. have proposed PS2MAN \cite{ps2man} for synthesizing the facial photo images from the sketch images by using the multi-adversarial networks with synthesized loss.
Other notable works are GAN based visible face synthesis (GAN-VFS) \cite{vfsgan}, Thermal-to-Visible GAN (TV-GAN) \cite{tvgan}, semantic-guided GAN (SG-GAN) \cite{MTVFISGGAN}, etc.

\begin{figure}[t]
\centering
\includegraphics[width=\columnwidth]{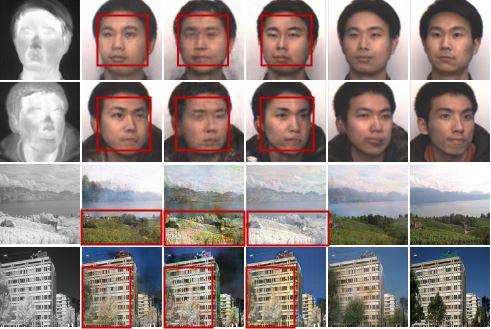}
\setlength{\belowcaptionskip}{-10pt} 
\caption{The efficacy of the proposed PCSGAN model. The sample images in $1^{st}$ column are taken from the WHU-IIP (i.e., thermal images in $1^{st}$ and $2^{nd}$ rows) and the RGB-NIR scene (i.e., NIR images in $3^{rd}$ and $4^{th}$ rows) datasets, respectively. The generated images using Pix2pix \cite{cGAN}, DualGAN \cite{dualGAN}, CycleGAN  \cite{CyclicGAN}, and proposed PCSGAN method are shown in $2^{nd}$, $3^{rd}$, $4^{th}$, and $5^{th}$ columns, respectively. The ground truth images in visible domain are shown in the last column. The rectangles in red color depict the artifacts, blurred and missing parts in the generated images by existing methods, which are overcome by the proposed PCSGAN method.}
\label{fig:qualititative}
\end{figure}

\begin{figure*}[t]
\centering
\includegraphics[width=\textwidth]{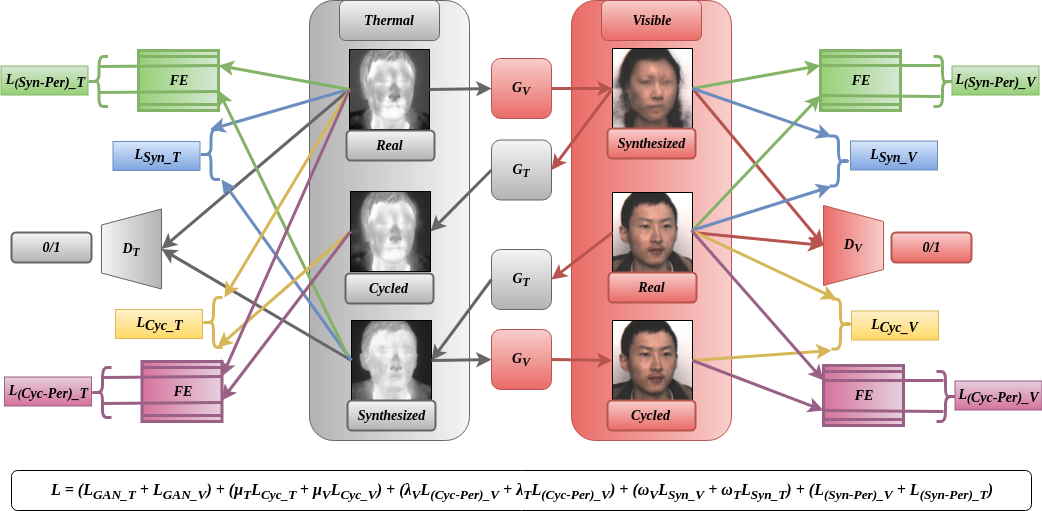}
\caption{The PCSGAN framework for Thermal-to-Visible image transformation. $G_V$ and $G_T$ are the generator networks for Thermal to Visible and Visible to Thermal transformations, respectively. $D_V$ and $D_T$ are the discriminators to distinguish between the Real\_Images (probability $1$) and Synthesized\_Images (probability $0$) in Visible and Thermal domains, respectively. ${L}_{LSGAN\_T}$ and ${L}_{LSGAN\_V}$ are the Adversarial losses, ${L}_{Cyc\_T}$ and ${L}_{Cyc\_V}$ are the Cycle-consistency losses, ${L}_{Syn\_T}$ and ${L}_{Syn\_V}$ are the Synthesized losses, ${L}_{(Cyc-Per)\_T}$ and ${L}_{(Cyc-Per)\_V}$ are the Cycled\_Perceptual losses and ${L}_{(Syn-Per)\_T}$ and ${L}_{(Syn-Per)\_V}$ are the Synthesized\_Perceptual losses. $FE$ is the feature extractor to extract the features from the images.}
\label{PCSGAN-Architecture}
\centering
\end{figure*}

It is pointed out from the above discussed literature that the vanilla version of GAN is not able to produce very realistic and artifact free images. The existing GAN based methods rely over the dedicated and specific losses in its objective function. Moreover, in order to improve the performance, some methods use the additional information computed in prior. Thus, there is a need to find the suitable objective/loss functions for the image-to-image transformation problem. 

A new method for Thermal (THM)/Near-Infrared (NIR) to Visible (VIS) image transformation called Perceptual Cyclic-Synthesized Generative Adversarial Networks (PCSGAN) is introduced in this paper. The PCSGAN method consists of two different networks, called the generator and the discriminator network like CycleGAN \cite{CyclicGAN}. 
It is like a two player mini-max game, where the discriminator network tries to maximize the given objective function by correctly differentiating between the real and the generated VIS images. Meanwhile, the generator network tries to minimize the same objective function by generating real looking VIS image to fool the discriminator network as depicted in Fig. \ref{fig:qualititative}. In addition to the adversarial losses proposed in GAN \cite{gan} and the pixel-wise similarity losses proposed in \cite{cGAN}, \cite{CyclicGAN}, we also use perceptual losses introduced in \cite{PLRTSTSR} for THM/NIR to VIS image synthesis.

The main contributions are summarized as follows:
\begin{itemize}
  \item A new method for Thermal/NIR to Visible image transformation called Perceptual Cyclic-Synthesized  Generative Adversarial Network (PCSGAN) is proposed.
  \item The PCSGAN utilizes two perceptual losses called the Cycled\_Perceptual loss and the Synthesized\_Perceptual loss in addition to the Adversarial and Pixel-wise losses. 
  \item The detailed experiments are conducted to show the improved performance of the proposed PCSGAN method over the WHU-IIP face and the RGB-NIR scene datasets. 
  \item Further, the ablation studies on losses are also conducted to verify the effectiveness of the added losses. 
  \end{itemize}
The rest of the paper is organized in following sections: the proposed PCSGAN method is described in Section \ref{Proposed method} along with the different loss functions and architecture; Section \ref{ExpSetup} describes the different datasets, evaluation metrics, and state-of-the-art methods; Section \ref{ResultsAndAnalysis} shows the experimental results with analysis. Section \ref{AblationStudy} conducts the ablation studies; and Section \ref{conclusion} concludes with future directions.

\section{Proposed Method}
\label{Proposed method}
In this section, the proposed PCSGAN framework along with the different loss functions is presented and the networks details are explained.

\begin{table*}
\caption{The relationship between the different loss functions used in the PCSGAN model and recent models, including GAN \cite{gan}, Pix2pix \cite{cGAN}, DualGAN\cite{dualGAN}, CycleGAN \cite{CyclicGAN}, PS2GAN \cite{ps2man}, and PAN \cite{PAN}, respectively. **DualGAN method objective function is similar to the CycleGAN method.}
\label{loss_comparison_table}
\centering
\begin{tabular}{c c c c c c c c }
\hline
\textbf{Losses} & \multicolumn{7}{ c }{\textbf{Methods}}  \\ 
\cline{2-8}
&  GAN \cite{gan} & Pix2pix \cite{cGAN} & **DualGAN \cite{dualGAN}  & CycleGAN \cite{CyclicGAN} & PS2GAN \cite{ps2man}  & PAN \cite{PAN}  & PCSGAN (Ours)   \\ 
 \hline 
 $\mathbf{\mathcal{L}_{LSGAN\_T}}$ & $\checkmark$ & $\checkmark$ & $\checkmark$ & $\checkmark$  & $\checkmark$ & $\checkmark$ & $\checkmark$   \\
$\mathbf{\mathcal{L}_{LSGAN\_V}}$ & $\checkmark$ & $\checkmark$ & $\checkmark$ & $\checkmark$ & $\checkmark$ & $\checkmark$& $\checkmark$  \\

$\mathbf{\mathcal{L}_{Cyc\_T}}$ &  &  &$\checkmark$  & $\checkmark$&  $\checkmark$& &$\checkmark$   \\

$\mathbf{\mathcal{L}_{Cyc\_V}}$&  &  &$\checkmark$  & $\checkmark$&  $\checkmark$& &$\checkmark$   \\

$\mathbf{\mathcal{L}_{(Cyc-Per)\_T}}$&  &  &  & & & &$\checkmark$  \\

$\mathbf{\mathcal{L}_{(Cyc-Per)\_V}}$&  &  &  & & & &$\checkmark$  \\

$\mathbf{\mathcal{L}_{Syn\_T}}$ &   & $\checkmark$ & & & $\checkmark$& & $\checkmark$\\ 

$\mathbf{\mathcal{L}_{Syn\_V}}$ &   & $\checkmark$ & & & $\checkmark$& & $\checkmark$\\

$\mathbf{\mathcal{L}_{(Syn-Per)\_T}}$ & &  &  & & & &$\checkmark$\\ 

$\mathbf{\mathcal{L}_{(Syn-Per)\_V}}$ & &  &  & & & &$\checkmark$\\

$\mathbf{\mathcal{L}_{PAL\_T}}$ & &  &  & & & $\checkmark$&\\ 

$\mathbf{\mathcal{L}_{PAL\_V}}$ & &  &  & & & $\checkmark$&\\

\hline
\end{tabular}
\end{table*}

\subsection{The PCSGAN Framework}
The PCSGAN framework as shown in the Fig. \ref{PCSGAN-Architecture} is used to transform the images from the source Thermal/NIR domain $T$ to the target Visible domain $V$. For ease of understanding, we consider Thermal to Visible transformation, the same explanation applies for NIR to Visible also. 
The PCSGAN framework consists two generator networks $G_{V}: T \rightarrow V$ and $G_{T}: V \rightarrow T$ to transform the images from Thermal to Visible and Visible to Thermal, respectively. These generator networks are trained adversarially by using two discriminator networks $D_V$ and $D_T$ in domains $V$ and $T$, respectively.  

For illustration, the first generator $G_V$ receives an input image, $Real\_T$ in domain $T$ and transform it into the Synthesized\_Image in domain $V$ as $Syn\_V$, i.e., $Syn\_V=G_V(Real\_T)$. In this context, the generator network $G_V$ is trained to generate the $Syn\_V$ which looks same as $Real\_V$ to fool the discriminator $D_V$, such that it should not be able to distinguish between $Syn\_V$ and $Real\_V$. Whereas, the  discriminator is $D_V$ trained to distinguish clearly between the $Real\_V$ as real and $Syn\_V$ as generated. 
On the other hand, the second generator $G_T$ transforms the Real Image from domain $V$ into the Synthesized Image in domain $T$ as, $Syn\_T=G_T(Real\_V)$. Here, the generator network $G_T$ is trained to generate the $Syn_T$ close to $Real\_T$ to fool the discriminator $D_T$ to think $Syn\_T$ as $Real\_T$. Whereas, the  discriminator $D_T$ distinguishes the $Real\_T$ as real and $Syn\_T$ as fake. Thus, the generator networks $G_V$ and $G_T$ are trained adversarially by competing with the discriminator networks $D_V$ and $D_T$, respectively. We use the adversarial loss functions ${L}_{LSGAN\_V}$ and ${L}_{LSGAN\_V}$ introduced in \cite{LSGAN} to train the combined generator and discriminator networks.

\begin{dmath}
\label{LSGANV}
{min}_{G_V}{max}_{D_V}\mathcal{L}_{LSGAN\_V}(G_{V}, D_V) =  \scriptstyle{E}_{V \sim P_{data}(V)} [(D_V(Real\_V)-1)^2]  + \scriptstyle{E}_{T \sim P_{data}(T)} [D_V(G_{V}(Real\_T))^2].
\end{dmath}

\begin{dmath}
\label{LSGANA}
{min}_{G_T}{max}_{D_T}\mathcal{L}_{LSGAN\_T}(G_{T}, D_T) =  \scriptstyle{E}_{T \sim P_{data}(T)} [(D_T(Real\_T)-1)^2]
+ \scriptstyle{E}_{V \sim P_{data}(V)} [D_T(G_{T}(Real\_V))^2].
\end{dmath}

where, ${L}_{LSGAN\_V}$ and ${L}_{LSGAN\_T}$ are the adversarial losses in the domains $T$ and $V$, respectively. The adversarial losses guide the generator to produce the images in the target domain.

\subsubsection{Pixel-wise Similarity Loss Functions} 
In this work, two pixel-wise similarity loss functions called Cycle-consistency loss introduced in \cite{CyclicGAN} and Synthesized loss introduced in \cite{ps2man} are included. The CycleGAN is originally proposed for unpaired datasets and in this case training the model with only adversarial losses leads to the mode collapse problem, where most of the images from a source domain mapped to the single image in the target domain. 

Cycle-consistency loss, calculated as the $L_1$ loss between the Real Images ($Real\_T$ and $Real\_V$) and Cycled Images ($Cyc\_T$ and $Cyc\_V$), is used to overcome from the above mentioned problem. The $Cyc\_T$ and $Cyc\_V$ are computed as,

\begin{equation}
\label{cycled_imagesT}
Cyc\_T=G_{T}(Syn\_V)=G_{T}(G_{V}(Real\_T))
\end{equation}
\begin{equation}
\label{cycled_imagesV}
Cyc\_V=G_{V}(Syn\_T)=G_{V}(G_{T}(Real\_V))
\end{equation}
The Cycle-consistency loss in domain $T$ and domain $V$ are represented as ${L}_{Cyc\_T}$ and ${L}_{Cyc\_V}$, respectively, and given as,
\begin{equation}
\label{LCYCT}
\mathcal{L}_{Cyc\_T}=\left \| Real\_T-Cyc\_T \right \|_1
\end{equation} 
\begin{equation}
\label{LCYCV}
\mathcal{L}_{Cyc\_V}=\left \| Real\_V-Cyc\_V \right \|_1
\end{equation}
where, ${L}_{Cyc\_T}$ is the Cycle-consistency loss calculated between the Real\_Image ($Real\_T$) and Cycled\_Image ($Cyc\_T$) in domain $T$ using the $L_1$ loss and ${L}_{Cyc\_V}$ is the Cycle-consistency loss calculated between the Real\_Image ($Real\_V$) and Cycled\_Image ($Cyc\_V$) in domain $V$ using the $L_1$ loss. In the THM to VIS image transformation, these Cycle-consistency losses act as an additional regularizers and help to learn the network parameters by reducing the artifacts in the produced images. 

Synthesized loss is introduced based on the observation that the task of the generator is not only to generate the synthesized images for fooling the discriminator, but also the generated images should look like realistic and closer to the target domain. This is not possible only with the Adversarial loss and Cycle-consistency loss. Synthesized loss is computed between the Real\_Images ($Real\_T$ and $Real\_V$) and the Synthesized Images ($Syn\_T$ and $Syn\_V$) in domain $T$ and $V$, respectively. The Synthesized losses in domain $T$ and $V$ are represented as ${L}_{Syn\_T}$ and ${L}_{Syn\_V}$ and defined as,
\begin{equation}
\label{LST}
\mathcal{L}_{Syn\_T}=\left \| Real\_T-Syn\_T \right \|_1
\end{equation}
where, ${L}_{Syn\_T}$ is the Synthesized loss calculated between the Real Image ($Real\_T$) and Synthesized Image ($Syn\_T$) in domain $T$ using the $L_1$ loss, and, 
\begin{equation}
\label{LSV}
\mathcal{L}_{Syn\_V}=\left \| Real\_V-Syn\_V \right \|_1
\end{equation}
where, ${L}_{Cyc\_V}$ is the Synthesized loss calculated  between the Real Image ($Real\_V$) and Synthesized Image ($Cyc\_V$) in domain $V$ using the $L_1$ loss. 

\subsubsection{Perceptual Similarity Loss Functions}
For Thermal/NIR to Visible image transformation, the above discussed pixel-wise losses help the generator to produce the images closer to the target domain from pixel-wise content perspective. However, both the Cycle-consistency loss and Synthesized loss fail to get perceptual information for human judgment on the quality of the images \cite{context_enc}. So, when only pixel-wise similarity losses are used for image transformation, the generated images generally suffer with the reduced sharpness and missing fine-details in the structures \cite{PAN}. To solve this problem, feature-based loss functions were introduced in \cite{PLRTSTSR} to provide additional constraints to enhance the quality of the transformed image. In this work, we also utilize the power of feature-based loss by adding two more additional perceptual losses, namely, Cycled Perceptual loss and Synthesized Perceptual loss.

The Cycled\_Perceptual loss is calculated by extracting the intermediate features with the help of pre-trained feature extractor network ($\phi$), like VGG-19 or ResNet-50. It is computed between the Real Images ($Real\_T$ and $Real\_V$) and the Cycled Images ($Cyc\_T$ and $Cyc\_V$) in domain $T$ and $V$, represented as ${L}_{(Cyc-Per)\_T}$ and ${L}_{(Cyc-Per)\_V}$, respectively. These losses are given as,
\begin{equation}
\begin{split}
\label{Per-Cyc_T}
\mathcal{L}_{(Cyc-Per)\_T} &=  \scriptstyle{E}_{t, v \sim P_{data}(t,v)} \left \| \phi_P(G_T(Real\_V)- \phi_P(Real\_T)\right\|_1.
\end{split}
\end{equation}
where, ${L}_{(Cyc-Per)\_T}$ is the Cycled Perceptual loss calculated as mean absolute error (MAE) by extracting features between the Real Image ($Real\_T$) and Cycled Image ($Cyc\_T$) in domain $T$, and,  
\begin{equation}
\begin{split}
\label{Cyc-Per_V}
\mathcal{L}_{(Cyc-Per)\_V} &= \scriptstyle{E}_{t, v \sim P_{data}(t,v)} \left \| \phi_P(G_V(Real\_T)- \phi_P(Real\_V)\right\|_1.
\end{split}
\end{equation}
where, ${L}_{(Cyc-Per)\_V}$ is the Cycled Perceptual loss calculated as mean absolute error (MAE) by extracting the features between the Real Image ($Real\_V$) and Cycled Image ($Cyc\_V$) in domain $V$. 

Synthesized Perceptual loss is similar to the Synthesized loss, instead of the pixel-wise loss it is computed using the feature loss. This loss is computed between the Real Images ($Real\_T$ and $Real\_V$) and Synthesized Images ($Syn\_T$ and $Syn\_V$) in domain $T$ and $V$, represented as ${L}_{(Syn-Per)\_T}$ and ${L}_{(Syn-Per)\_V}$, respectively. These losses are computed as,
\begin{equation}
\begin{split}
\label{Syn-Per-T}
\mathcal{L}_{(Syn-Per)\_T} &=  \scriptstyle{E}_{t, v \sim P_{data}(t,v)} \left \| \phi_P(G_T(Real\_V)- \phi_P(Real\_T)\right\|_1.
\end{split}
\end{equation}
where, ${L}_{(Syn-Per)\_T}$ is the Synthesized Perceptual loss computed as the mean absolute error (MAE) by extracting the features from the Real Image ($Real\_T$) and Synthesized\_Image ($Cyc\_T$) in domain $T$, and, 
\begin{equation}
\begin{split}
\label{Syn-Per-V}
\mathcal{L}_{(Syn-Per)\_V} &=  \scriptstyle{E}_{t, v \sim P_{data}(t,v)} \left \| \phi_P(G_V(Real\_T)- \phi_P(Real\_V)\right\|_1.
\end{split}
\end{equation}
where, ${L}_{(Syn-Per)\_V}$ is the Synthesized Perceptual loss computed as the mean absolute error (MAE) by extracting the features from the Real Image ($Real\_V$) and the Synthesized Image ($Syn\_V$) in domain $V$.

\subsubsection{Final Objective Function}
The final objective function of the PCSGAN framework consists of the Adversarial losses, Cycle-consistency losses, Synthesized losses, Cycled Perceptual losses and Synthesized Perceptual losses. It is given as,
\begin{equation}
\label{LFINAL}
\begin{split}
\mathcal{L}_(G_{T}, G_{V}, D_T,
D_V)=\mathcal{L}_{LSGAN\_T}+\mathcal{L}_{LSGAN\_V}  +\lambda_T\mathcal{L}_{Cyc\_T}  +\lambda_V{\mathcal{L}_{Cyc\_V}} \\ +\mu_T\mathcal{L}_{Syn\_T}  +\mu_V{\mathcal{L}_{Syn\_V}} + \omega_T\mathcal{L}_{(Per-Cyc)\_T}  + \omega_V\mathcal{L}_{(Per-Cyc)\_V} \\ + \psi_T\mathcal{L}_{(Syn-Per)\_T} +{\psi_V\mathcal{L}_{(Syn-Per)\_V}}.
\end{split}
\end{equation}
where, ${L}_{LSGAN\_T}$ and ${L}_{LSGAN\_V}$ are the Adversarial losses, ${L}_{Cyc\_T}$ and ${L}_{Cyc\_V}$ are the Cycle-consistency losses, ${L}_{Syn\_T}$ and ${L}_{Syn\_V}$ are the Synthesized losses, ${L}_{(Cyc-Per)\_T}$ and ${L}_{(Cyc-Per)\_V}$ are the Cycled\_Perceptual losses and ${L}_{(Syn-Per)\_T}$ and ${L}_{(Syn-Per)\_V}$ are the Synthesized\_Perceptual losses. 
The weights are set empirically for the different losses used in the final objective function which are as follows: $\lambda_T=10$, $\lambda_V=10$, $\mu_T=15$, $\mu_V=15$, $\omega_T=1$, $\omega_V=1$, $\psi_T=1$ and $\psi_V=1$.
The relation between the different loss functions used in the proposed PCSGAN method and the state-of-the-art methods is summarized in Table \ref{loss_comparison_table}. The losses ${L}_{PAL\_T}$ and ${L}_{PAL\_V}$ shown in the Table \ref{loss_comparison_table} are the Perceptual Adversarial Losses introduced in the PAN \cite{PAN} method calculated in domains $T$ and $V$, respectively.

\subsection{Implementation Details}
As we can see from the Fig. \ref{PCSGAN-Architecture} that the PCSGAN framework contains two generator networks ($G_T$ and $G_V$) and two discriminator networks ($D_T$ and $D_V$). One generator and one discriminator are in the source domain $T$, whereas another generator and another discriminator are in the target domain $V$.
\subsubsection{Generator Network}
Similar to the CycleGAN \cite{cGAN}, we adopt the network with $9$ residual blocks from Jhonson et al. \cite{PLRTSTSR} as our generators networks. The generator network consists of: $3$ convolution layers $C7S1F64$, $C3S2F128$ and $C3S2F256$, followed by $9$ residual blocks $RB256$ and $2$ transpose convolution layers $DC3S2F128$, $DC3S2F64$ and one final convolution layer  $C7S1F3$.
A convolution with $x$ number of filters of size $y \times y$ with stride $z$ is represented by $CySzFx$. A residual block consisting of two Conv layers with $x$ filters is denoted by $RBx$. A deconvolution layer with $x$ filters of size $y \times y$ having stride $z$ is denoted by $DCySzFx$. Instance Norm is also used second convolution and all deconvolution layers. The ReLU activation function is used.
All the input images given to the PCSGAN method for training are resized to $256\times256$.

\subsubsection{Discriminator Network}
A $70\times70$ PatchGAN is used in the discriminator network similar to \cite{cGAN}. The layers of discriminator network are as follows: $4$ hidden layers $C4S2F64$, $C4S2F128$, $C4S2F256$ and $C4S2F512$. The Instance Norm is used all layers except first convolution layer. The final one-dimensional output is computed by a $4\times4$ convolutional layer having stride $1$. The activation function used in Leaky ReLU having slope $0.2$.

\subsubsection{Training Details}
In this work, as the proposed PCSGAN method is an enhancement of the CycleGAN method \cite{cGAN}, the training details are same as in the CycleGAN. The input images of fixed size $256\times256$ are given to the network as mentioned in \cite{cGAN}. The generator network with $9$ residual blocks, is best suitable for this size. We train the generator and the discriminator networks for $200$ epochs with only one sample per batch. Initially, we use $0.0002$ learning rate for $100$ epochs and then linearly decaying to $0$ for the next $100$ epochs. The networks are initialized with the Gaussian distribution having $0$ mean and $0.02$ standard deviation. The Adam \cite{adam} optimizer with momentum term $\beta_1$ as $0.9$ is used for optimizing the network.

\section{Experimental Setup}
\label{ExpSetup}
\subsection{Data Sets}
In this experiment, WHU-IIP Thermal and Visible face dataset and RGB-NIR Near-Infrared to Visible scene dataset are used. The WHU-IIP\footnote{http://iip.whu.edu.cn/projects/IR2Vis\_dataset.html\label{WHU_IIP}} face dataset contains a total of $792$ paired thermal and visible facial images taken from $33$ individuals \cite{TVFITGAN}. We use $552$ samples from $23$ individuals in training set and $240$ samples from $10$ individuals in test set. The RGB-NIR\footnote{https://ivrl.epfl.ch/research-2/research-downloads/supplementary\_material-cvpr11-index-html/} scene dataset consists the samples from $9$ classes with $477$ images in total. The image pairs are captured in both Near-infrared (NIR) as well as Visible (RGB) domains. The training and testing sets contain $387$ and $90$ samples, respectively.

\subsection{Evaluation Metrics}
In order show the improved outcome of the proposed PCSGAN method, we use the qualitative as well as the quantitative measures. Baseline image quality evaluation metrics, like Structural Similarity Index (SSIM) \cite{SSIM}, Peak Signal to Noise Ratio (PSNR), Mean Square Error (MSE), Learned Perceptual Image Patch Similarity (LPIPS) \cite{LPIPS}, and Multiscale Structural Similarity Index (MS-SSIM) \cite{MSSSIM} are used as the quantitative measures.

\subsection{State-Of-The-Art Compared Methods}
The four state-of-the-art Image-to-image transformation methods are used for comparison purpose, namely, Pix2pix, DualGAN, CycleGAN and PS2GAN. In order to have a fair comparison, we evaluate all the methods in paired setting only. 
\subsubsection{Pix2pix}
The publicly available code from Pix2pix\footnote{https://github.com/phillipi/pix2pix \label{Pix2pix}} \cite{cGAN} is used with the default settings.
\subsubsection{DualGAN}
We use DualGAN with the default settings as per the code available\footnote{https://github.com/duxingren14/DualGAN}\cite{dualGAN}.
\subsubsection{CycleGAN}
We use CycleGAN with the default settings as per the code available\footnote{https://github.com/junyanz/pytorch-CycleGAN-and-Pix2pix}\cite{CyclicGAN}.
\subsubsection{PS2GAN}
For this method, the original code taken from the authors consists of multiple adversarial networks \cite{ps2man}. It is originally proposed for generating the result images by calculating the losses at different resolutions of the given input images. For the fair comparison with the remaining state-of-the-art methods, we implement the PS2MAN with single adversarial networks, i.e., PS2GAN by adding Synthesized loss to the CycleGAN\cite{CyclicGAN} with other existing losses.
\subsubsection{PAN}
The code is taken from the PAN \footnote{https://github.com/DLHacks/pix2pix\_PAN}\cite{PAN}, the same settings mentioned in the original paper are used for the experiment.

\begin{table}[!t]
\caption{Quantitative evaluation of the results compared between the PCSGAN method and the state-of-the-art methods using the SSIM, MSE, PSNR, LPIPS, and MSSIM scores over the WHU-IIP face dataset.}
\label{whu-iip_main_table}
\centering
\scalebox{0.87}{
\begin{tabular}{|c|c|c|c|c|c|c|}
\hline
\textbf{Methods} & \multicolumn{5}{ c |}{\textbf{Metrics}}  \\ 
\cline{2-6}
& SSIM & MSE & PSNR & LPIPS & MSSIM\\
 \hline
Pix2pix \cite{cGAN} & $0.7555$ & $74.6082$ & $29.4587$ & $0.089$  & $0.7624$\\
\hline
 DualGAN \cite{dualGAN} & $0.7638$ & $75.4379$ & $29.4201$ & $0.099$  & $0.7989$\\
\hline
 CycleGAN \cite{CyclicGAN} & $0.7648$ & $76.1482$ & $29.351$ & $0.088$ & $0.7687$
\\
\hline
 PS2GAN \cite{ps2man} & $0.8087$ & $67.869$ & $29.9676$ & $0.064$ & $0.8242$ \\

\hline
PAN \cite{PAN} &$0.8125$  &$69.0331$  & $29.84$ & $0.069$ & $0.8281$ \\

\hline
 PCSGAN (Ours)& $\mathbf{0.8275}$ & $\mathbf{64.6442}$ & $\mathbf{30.1686}$ & $\mathbf{0.059}$ & $\mathbf{0.8411}$\\
\hline

\end{tabular}
}
\end{table}

\section{Experimental Results and Analysis}
\label{ResultsAndAnalysis}
In the experiments, the proposed PCSGAN method is evaluated in terms of the quantitative, qualitative and complexity measures.

\subsection{Quantitative Results}
The PCSGAN method is quantitatively evaluated by using five widely used image quality assessment metrics, namely SSIM, MSE, PSNR, LPIPS and MS-SSIM. The PCSGAN method clearly shows the improved performance over the state-of-the-art methods as shown in the Table \ref{whu-iip_main_table} and \ref{rgb-nir_main_table} over the WHU-IIP face dataset and RGB-NIR scene dataset, respectively. 
The PCSGAN shows an improvement over {Pix2pix \cite{cGAN}, DualGAN \cite{dualGAN}, CycleGAN \cite{CyclicGAN}, PS2GAN \cite{ps2man}} and PAN \cite{PAN} with 
\begin{itemize}
    \item an increment of \{$9.53$, $8.34$, $8.2$, $2.32$ and $1.84$\}, \{$2.41$, $2.54$, $2.79$, $0.67$ and $1.1$\} and \{$10.32$, $5.28$, $9.42$, $2.05$ and $1.56$\} in \% in terms of the SSIM, the PSNR and the MS-SSIM scores, respectively, on the WHU-IIP face dataset,
    \item a reduction of \{$13.36$, $14.31$, $15.11$, $4.75$ and $6.35$\} and \{$33.71$, $40.41$, $32.96$, $7.81$ and $14.49$\} in \% in terms of the MSE and the LPIPS score, respectively on the WHU-IIP face dataset, 
    \item an increment of \{$19.97$, $5587.3$, $18.19$, $8.54$ and $35.54$\} and \{$0.22$, $1.54$, $0.33$, $0.2$ and $0.82$\} in \% in terms of the SSIM and the PSNR scores, respectively, on the RGB-NIR scene dataset, and
    \item a reduction of \{$1.3$, $8.9$, $2.39$, $1.37$ and $5.03$\}, \{$27.69$, $56.62$, $31.91$, $14.67$ and $38.16$\} in \% in terms of the MSE and the LPIPS score, respectively on the RGB-NIR scene dataset.
\end{itemize}

These quantitative results confirm the superiority of the proposed PCSGAN method compared to the state-of-the-art methods.

\subsection{Qualitative Results}
\begin{figure*}[!t]
\centering
\includegraphics[width=\textwidth]{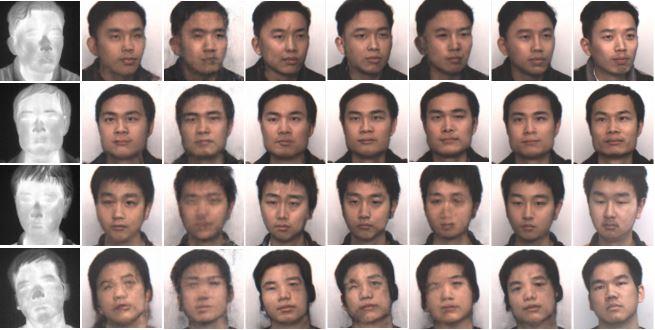}
\vspace{-5mm}
\caption{The qualitative comparison of results over WHU\_IIP dataset. From left-to-right input image, images generated by Pix2pix, DualGAN, CycleGAN, PS2GAN, PAN, PCSGAN and ground truth image, respectively.}
\label{whu-iip_main_fig}
\end{figure*}

\begin{figure*}[!t]
\centering
\includegraphics[width=\textwidth]{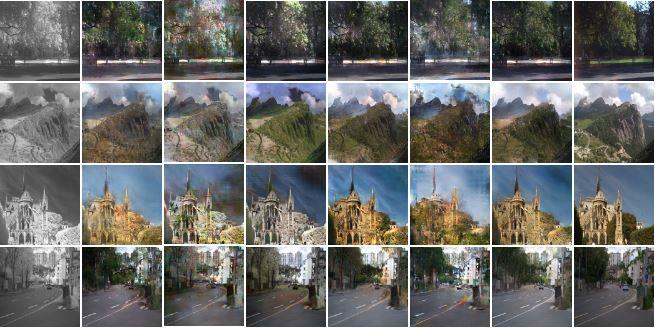}
\vspace{-5mm}
\caption{The qualitative comparison of results over RGB\_NIR scene dataset. From left-to-right input image, images generated by Pix2pix, DualGAN, CycleGAN, PS2GAN, PAN, PCSGAN and ground truth image, respectively.}
\label{rgb-nir_main_fig}
\end{figure*}

The qualitative results are also analyzed in this experiment to better understand the visual quality of the generated images in Fig. \ref{whu-iip_main_fig} on WHU-IIP face dataset and Fig. \ref{rgb-nir_main_fig} on RGB-NIR scene dataset. 
The qualitative comparison of results over the WHU-IIP face dataset is described as follows:
\begin{itemize}
    \item It can be observed that the quality and fine details (i.e., facial attributes) of the resulting face images generated by the proposed PCSGAN method are comparatively better than the state-of-the-art methods as shown in Fig. \ref{whu-iip_main_fig}.
    \item In particular, the $3^{rd}$ and $4^{th}$ rows in Fig. \ref{whu-iip_main_fig} show that the proposed PCSGAN method generated result images are much closer (more fine facial attribute details and less blurriness) to the ground truth images.
    \item It can also be observed that the state-of-the art methods fail (see facial attributes and blurriness in the images) to generate the result images closer to the ground truth images.  
\end{itemize}

\begin{table}[!t]
\caption{Quantitative evaluation of the results compared between the PCSGAN method and the state-of-the-art methods using the SSIM, MSE, PSNR, and LPIPS scores over the RGB-NIR scene dataset.}
\label{rgb-nir_main_table}
\centering
\scalebox{0.87}{
\begin{tabular}{|c|c|c|c|c|c|}
\hline
\textbf{Methods} & \multicolumn{4}{ c |}{\textbf{Metrics}}  \\ 
\cline{2-5}
& SSIM & MSE & PSNR & LPIPS \\
 \hline
Pix2pix \cite{cGAN} & $0.5763$ & $97.3463$ & $28.268$ & $0.177$  \\
\hline
 DualGAN \cite{dualGAN} & $-0.0126$ & $105.4514$ & $27.9019$ & $0.295$   \\
\hline
 CycleGAN \cite{CyclicGAN}& $0.585$ & $98.4225$ & $28.2377$ & $0.188$
\\
\hline
PS2GAN \cite{ps2man} & $0.637$ & $97.4073$ & $28.2753$ & $0.15$\\
\hline

PAN \cite{PAN} & $0.5101$ & $101.1342$  &$28.0978$  & $0.207$  \\
\hline

PCSGAN (Ours) & $\mathbf{0.6914}$ & $\mathbf{96.0744}$ & $\mathbf{28.3305}$ & $\mathbf{0.128}$ \\
\hline
\end{tabular}
}
\end{table}

In a similar way, the qualitative comparison of results over the RGB-NIR scene dataset is described as follows:
\begin{itemize}
    \item From the Fig. \ref{rgb-nir_main_fig}, it can be observed that the quality and fine details (i.e., color and texture) of the resulting scene images generated by the proposed PCSGAN method are comparatively better than the state-of-the-art methods.
    \item In particular from $2^{nd}$, $3^{rd}$ and $4^{th}$ columns of the Fig. \ref{rgb-nir_main_fig}, it can be observed that Pix2pix, DulGAN and CycleGAN methods completely fail to predict the color and texture of the target domain images. Whereas, it can be seen from $5^{th}$ and $6^{th}$ columns that the PS2GAN and PAN methods are able to generate the images, somewhat closer to the ground truth images and still can be improved further.
    \item The PCSGAN method generates higher quality and more realistic images compared to the state-of-the-art image-to-image transformation methods and same can be observed from $7^{th}$ column.  
\end{itemize}
The qualitative comparisons of the results over the WHU-IIP face and NIR-RGB scene datasets clearly show that the proposed PCSGAN method achieves much better results than the state-of-the-art image-to-image transformation methods.

\subsection{Complexity Analysis}
\label{CC}

\begin{table*}[t]
\caption{The complexity comparison of the proposed PCSGAN model with Pix2pix, DualGAN, CycleGAN, PS2GAN and PAN models over WHU-IIP face dataset.}
\label{complexities_comparison_table}
\centering
\scalebox{0.93}{
\begin{tabular}{|c|c|c|c|c|c|c|c|}
\hline
\textbf{Criteria} & \multicolumn{6}{ c |}{\textbf{Methods}}  \\ 
\cline{2-7}
& Pix2Pix & DualGAN & CycleGAN & PS2GAN & PAN & PCSGAN\\
 \hline
  \#Generators  & $1$ & $2$ & $2$ & $2$  & $1$ & $2$\\
 \hline
 \#Discriminators & $1$ & $2$ & $2$ & $2$  & $1$ & $2$\\
\hline
\#Gen\_Parameters & $11,378,179$ & $84,612,352$ & $11,378,179$ & $11,378,179$ & $54,413,955$ &$11,378,179$\\
\hline
 \#Dis\_Parameters & $2,767,809$ & $33,687,041$ & $2,764,737$ & $2,764,737$ & $2,768,705$ & $2,764,737$ \\

\hline
 Residual Network & $No$ & $No$ & $No$ & $No$ & $No$ & $Yes$  \\

\hline
 Memory& $2264\ MB$ & $10924\ MB$ & $3626\ MB$ & $3642\ MB$ & $2554\ MB$ & $4626\ MB$ \\
\hline
Training Time & $2:56\ hrs$ & $30:44\ hrs$ & $8:12\ hrs$ & $8:23\ hrs$ & $2:00\ hrs$ & $11:50\ hrs$\\
\hline
Testing Time& $45\ secs$ & $52\ secs$ & $81\ secs$ & $84\ secs$ & $48\ secs$ & $92\ secs$ \\
\hline
\end{tabular}
}
\end{table*}

The computational complexity of the proposed method and the state-of-the art methods are also compared. For this experiment, we use Titan X \(Pascal\) 12GB GPU, Intel(R) Core(TM) i7-7700 @ 3.60GHz CPU, and 64GB RAM based computer system. Pytorch 1.2.0 is used for all models except DualGAN for which Tensorflow 1.15.0 is used. The $552$ training images and $240$ test images from WHU_IIP face dataset are used in this experiment. The complexity is reported in Table \ref{complexities_comparison_table} in terms of the no. of Generators and parameters, no. of Discriminators and parameters, residual based on not, memory, training time and test time. The training time is computed only once, whereas the test time is calculated as the average of $10$ runs.
\\
The important points in the analysis are as follows:
\begin{itemize}
    \item The proposed PCSGAN uses two Generator and two Discriminator networks similar to DualGAN, CycleGAN and PAN.
    \item It is also evident that the no. of parameters in the Generator and Discriminator networks of the proposed PCSGAN are same as CycleGAN and PS2GAN which can be seen as the lowest among all. 
    \item The proposed PCSGAN method consists of an additional residual network for calculating the perceptual loss, which is not present in other compared state-of-the-art methods.  
    \item The proposed PCSGAN method is memory expensive due to the additional residual network used. However, it is more efficient than the DualGAN model. The Pix2Pix and PAN are memory efficient due to the presence of only one Generator and only one Discriminator network in both models. The DualGAN needs more memory due to more number of parameters.
    \item The training time is generally dependent upon the number of parameters and convergence of the model. The training time of the proposed PCSGAN model is increased due to the inclusion of the additional residual network.
    \item The Discriminator networks do not play any role during the test time in the GAN models. The test time of the proposed PCSGAN model is comparable with the CycleGAN and PS2GAN models. The test time using the DualGAN model is less as it is computed in Tensorflow, whereas the test time using other models is computed in Pytorch. 
\end{itemize}

\begin{figure*}[t]
\centering
\includegraphics[width=\textwidth]{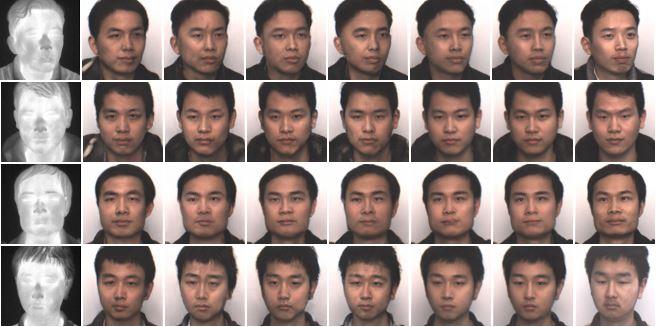}
\caption{The generated images using different different losses in PCSGAN framework over the WHU\_IIP face dataset. The first and last columns represent the input and target images, respectively. The $2^{nd}$ to $7^{th}$ columns, from left to right, shows the generated images using AL, AL++CL, AL+CL+CPL, AL+CL+SL, AL+CL+SL+SPL, and AL+CL+CPL+SL+SPL losses, respectively.}
\label{whu-iip_ablation_fig}
\end{figure*}

\begin{figure*}[t]
\centering
\includegraphics[width=\textwidth]{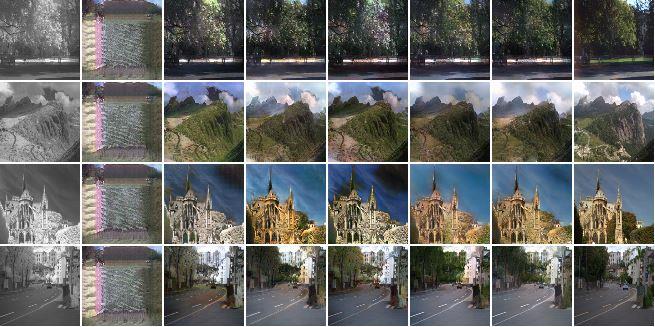}
\caption{The generated images using different different losses in PCSGAN framework over the RGB\_NIR scene dataset. The columns are similar to Fig. \ref{whu-iip_ablation_fig}.}
\label{rgb-nir_ablation_fig}
\end{figure*}

\begin{table*}[t]
\caption{Ablation study over different loss functions used in the proposed PCSGAN over WHU-IIP face dataset.}
\label{whu-iip_ablation_table}
\centering
\scalebox{0.93}{
\begin{tabular}{|c|c|c|c|c|c|c|}
\hline
\textbf{Loss Functions} & \multicolumn{5}{ c |}{\textbf{Metrics}}  \\ 
\cline{2-6}
& SSIM & MSE & PSNR & LPIPS & MSSIM\\
 \hline
  AL  & $0.7555$ & $74.6082$ & $29.4587$ & $0.089$  & $0.7624$\\
 \hline
 AL+CL & $0.7648$ & $76.1482$ & $29.351$ & $0.088$ & $0.7687$
\\
\hline
AL+CL+CPL & $0.7461$ & $77.6016$ & $29.264$ & $0.096$ & $0.7456$\\
\hline
 AL+CL+SL & $0.8087$ & $67.869$ & $29.9676$ & $0.064$ & $0.8242$ \\

\hline
 AL+CL+SL+SPL & $0.8212$ & $68.029$ & $29.9296$ & $0.065$ & $0.8324$ \\

\hline
 AL+CL+CPL+SL+SPL& $\mathbf{0.8275}$ & $\mathbf{64.6442}$ & $\mathbf{30.1686}$ & $\mathbf{0.059}$ & $\mathbf{0.8411}$\\
\hline
\end{tabular}
}
\end{table*}

\section{Ablation Study}
\label{AblationStudy}
In this paper, we conduct an ablation study using different loss functions, namely, Adversarial losses, pixel-wise losses and perceptual losses. It is dedicated to better understand the impact of the newly added perceptual loss functions. For simplicity, we label the Adversarial Loss as $AL$, Cycle-consistency Loss as $CL$, Synthesized Loss $SL$, Cycle-Perceptual Loss as $CPL$ and Synthesized-Perceptual Loss as $SPL$.
The ablation study is performed over the WHU-IIP face and the RGB-NIR scene datasets in terms of both the quantitative measures (summarized in the Table \ref{whu-iip_ablation_table} and Table \ref{rgb-nir_ablation_table}) and the qualitative measures (illustrated in the Fig. \ref{whu-iip_ablation_fig} and Fig. \ref{rgb-nir_ablation_fig}), respectively. The observations from this ablation study are described as follows:
\subsection{AL}
In this setting, we use only the Adversarial losses mentioned in Equation \ref{LSGANV} and \ref{LSGANA} as the objective function for image transformation over the WHU-IIP face dataset and the RGB-NIR scene dataset.
Over the WHU-IIP dataset, the generated images suffer with severe artifacts and lack of facial attribute information. However, over the RGB-NIR dataset, the resulting images suffer with the mode collapse problem and completely fail to generate the images. Thus, the Adversarial losses alone, are unable to generate the good quality realistic images and lead to a high domain discrepancy gap. 

\begin{table}[t]
\caption{Ablation study over different loss functions used in the proposed PCSGAN over RGB-NIR scene dataset.}
\label{rgb-nir_ablation_table}
\centering
\scalebox{0.9}{
\begin{tabular}{|c|c|c|c|c|c|}
\hline
\textbf{Loss Functions} & \multicolumn{4}{ c |}{\textbf{Metrics}}  \\ 
\cline{2-5}
& SSIM & MSE & PSNR & LPIPS \\
 \hline
AL & $0.153$ & $104.8787$ & $27.9255$ & $0.403$
\\
\hline
 AL+CL & $0.585$ & $98.4225$ & $28.2377$ & $0.188$
\\
\hline
AL+CL+CPL & $0.6428$ & $98.7429$ & $28.2153$ & $0.157$ \\
\hline
AL+CL+SL & $0.637$ & $97.4073$ & $28.2753$ & $0.15$\\
\hline
AL+CL+SL+SPL & $0.684$ & $97.5858$ & $28.2732$ & $0.133$\\
\hline
AL+CL+CPL+SL+SPL & $\mathbf{0.6914}$ & $\mathbf{96.0744}$ & $\mathbf{28.3305}$ & $\mathbf{0.128}$ \\
\hline
\end{tabular}
}
\end{table}

\subsection{AL+CL}
In this setting, we use Cycle-consistency losses mentioned in Equation \ref{LCYCT} and \ref{LCYCV} along with the Adversarial losses as the objective function. From the results obtained, it can be seen that the Cycle-consistency loss help to overcome the mode collapse problem over the RGB-NIR scene dataset. Whereas, the generated images still suffer from the color disparity and lack of fine details over both the WHU-IIP and RGB-NIR scene datasets.
\subsection{AL+CL+CPL}
In this setting, we use Cycled\_Perceptual losses mentioned in Equation \ref{Per-Cyc_T} and \ref{Cyc-Per_V} along with the Adversarial losses and Cycle-Consistency losses as the objective function. From the results, it can be observed that the Cycled\_Perceptual losses help to generate the images with finer details and proper color judgment. 
\subsection{AL+CL+SL}
In this setting, we use Synthesized losses mentioned in Equation \ref{LST} and \ref{LSV} along with the Cycle-consistency losses and the Adversarial losses as the objective function. It is noticed from the results that synthesized losses boost the generator to produce the images not only to fool the discriminator, but also to look closer to the target domain. Although, the generated images are closer to the target domain images, still they suffer from lack of fine details in terms of the color, texture and shape. 
\subsection{AL+CL+SL+SPL}
In this setting, we use Synthesized\_Perceptual losses mentioned in Equation \ref{Syn-Per-T} and \ref{Syn-Per-V} along with the Synthesized\_losses, Cycle-consistency losses and Adversarial losses as the objective function. The results after adding the Synthesized\_Perceptual losses, preserve the fine details and generate images closer to the target domain. Still, there is a scope to decrease the domain discrepancy gap by adding Cycled\_Perceptual loss.
\subsection{AL+CL+CPL+SL+SPL}
For the proposed PCSGAN, we combine all the above mentioned losses, namely, Adversarial losses, Cycle-consistency losses, Cycled\_Perceptual losses, Synthesized losses and Synthesized\_Perceptual losses as the objective function. From the results, it is clear that the images generated by the proposed PCSGAN are more realistic with finer details and negligible artifacts when compared to the remaining above mentioned settings.

We also conducted an experiment by adding Perceptual Adversarial loss introduced in PAN \cite{PAN} to our proposed PCSGAN method, where we did not find much increase in the quality of the generated images. 

From this ablation study, it is pointed out that the Adversarial losses help to generate the images in the target domains, but the generated images suffer from artifacts, blurred portions, lack of fine details and sometimes from the mode collapse problem. The pixel-wise similarity loss functions, namely, Cycle-consistency losses and Synthesized losses help to generate images closer to the target domain and to reduce the artifacts. The perceptual losses, namely, Cycled\_Perceptual and Synthesized\_Perceptual losses force the network to generate the images with semantic and finer details in terms of the consistency in color, texture and shape for different regions.

Followings are the results summary:
\begin{enumerate}
    \item The proposed PCSGAN has shown the outstanding performance over the problem of Thermal/NIR to Visible image transformation. 
    \item The proposed PCSGAN outperforms the existing GAN methods such as Pix2Pix, DualGAN, CycleGAN, PS2GAN and PAN in terms of the different measures such as SSIM, MSE, PSNR, and LPIPS.
    \item The proposed PCSGAN generates the images with better quality and fine details.
    \item The importance of different losses used in the proposed PCSGAN is observed using the ablation study with different combination of losses in objective function.
\end{enumerate}

\section{Conclusion and Future Work}
\label{conclusion}

In this paper, we present a new PCSGAN model for transforming the images from Thermal/NIR domain to Visible domain. The proposed PCSGAN method uses the perceptual losses in addition to the adversarial and pixel-wise losses which are generally used by the recent state-of-the-art image-to-image transformation methods. The quality of the generated images has been greatly improved in terms of the finer details, reduced artifacts and semantics after addition of the perceptual losses, namely, Cycled\_Perceptual losses and Synthesized\_Perceptual losses. The same is observed through both the quantitative and qualitative measures. The proposed method outperforms the other state-of-the-art compared methods for Thermal-Visible and NIR-Visible transformation problems. The improved results are observed using the SSIM, PSNR, MSE and LPIPS performance measures. The ablation study confirms the the suitability and relevance of the added perceptual losses which boost the quality of the generated images.

The proposed approach can be applied to many other paired image-to-image transformation based applications like sketch-to-photo transformation, labels-to-facades transformation, aerial-to-maps transformation, night-to-day image translation, image enhancement, image colorization, image synthesis, etc.
The future work includes the extension over more than two modalities. Moreover, we also want to extend this work to find the effective losses and constraints for image transformation between the heterogeneous datasets, such as visible, sketch, thermal and NIR. The current GANs are heavy weight models and not suitable for the mobile devices. It is also one of the future directions to develop the lightweight GAN models. Another future research direction is to learn the GAN architectures automatically using the neural architecture search.

\section*{ACKNOWLEDGEMENT}
\label{acknowledgment}
We are thankful to NVIDIA Corporation for donating us the NVIDIA GeForce Titan X Pascal 12GB GPUs which is used in this research.

\bibliographystyle{IEEEtran}
\bibliography{References}

\begin{thebibliography}{10}
\providecommand{\url}[1]{#1}
\csname url@samestyle\endcsname
\providecommand{\newblock}{\relax}
\providecommand{\bibinfo}[2]{#2}
\providecommand{\BIBentrySTDinterwordspacing}{\spaceskip=0pt\relax}
\providecommand{\BIBentryALTinterwordstretchfactor}{4}
\providecommand{\BIBentryALTinterwordspacing}{\spaceskip=\fontdimen2\font plus
\BIBentryALTinterwordstretchfactor\fontdimen3\font minus
  \fontdimen4\font\relax}
\providecommand{\BIBforeignlanguage}[2]{{%
\expandafter\ifx\csname l@#1\endcsname\relax
\typeout{** WARNING: IEEEtran.bst: No hyphenation pattern has been}%
\typeout{** loaded for the language `#1'. Using the pattern for}%
\typeout{** the default language instead.}%
\else
\language=\csname l@#1\endcsname
\fi
#2}}
\providecommand{\BIBdecl}{\relax}
\BIBdecl

\bibitem{Applications}
S.~Hu, N.~Short, B.~S. Riggan, M.~Chasse, and M.~S. Sarfraz, ``Heterogeneous
  face recognition: recent advances in infrared-to-visible matching,'' in
  \emph{2017 12th IEEE International Conference on Automatic Face \& Gesture
  Recognition (FG 2017)}.\hskip 1em plus 0.5em minus 0.4em\relax IEEE, 2017,
  pp. 883--890.

\bibitem{gan}
I.~Goodfellow, J.~Pouget-Abadie, M.~Mirza, B.~Xu, D.~Warde-Farley, S.~Ozair,
  A.~Courville, and Y.~Bengio, ``Generative adversarial nets,'' in
  \emph{Advances in neural information processing systems}, 2014, pp.
  2672--2680.

\bibitem{cGAN}
P.~Isola, J.-Y. Zhu, T.~Zhou, and A.~A. Efros, ``Image-to-image translation
  with conditional adversarial networks,'' in \emph{IEEE Conference on Computer
  Vision and Pattern Recognition}, 2017, pp. 5967--5976.

\bibitem{AIMIM}
K.~Armanious, Y.~Mecky, S.~Gatidis, and B.~Yang, ``Adversarial inpainting of
  medical image modalities,'' in \emph{ICASSP 2019-2019 IEEE International
  Conference on Acoustics, Speech and Signal Processing (ICASSP)}.\hskip 1em
  plus 0.5em minus 0.4em\relax IEEE, 2019, pp. 3267--3271.

\bibitem{DLISR}
T.~Guo, H.~S. Mousavi, and V.~Monga, ``Deep learning based image
  super-resolution with coupled backpropagation,'' in \emph{IEEE Global
  Conference on Signal and Information Processing}, 2016, pp. 237--241.

\bibitem{SISRDL}
J.~Chen, X.~He, H.~Chen, Q.~Teng, and L.~Qing, ``Single image super-resolution
  based on deep learning and gradient transformation,'' in \emph{IEEE
  International Conference on Signal Processing}, 2016, pp. 663--667.

\bibitem{PLRTSTSR}
J.~Johnson, A.~Alahi, and L.~Fei-Fei, ``Perceptual losses for real-time style
  transfer and super-resolution,'' in \emph{European Conference on Computer
  Vision}, 2016, pp. 694--711.

\bibitem{PRSISRGAN}
C.~Ledig, L.~Theis, F.~Husz{\'a}r, J.~Caballero, A.~Cunningham, A.~Acosta,
  A.~P. Aitken, A.~Tejani, J.~Totz, Z.~Wang \emph{et~al.}, ``Photo-realistic
  single image super-resolution using a generative adversarial network.'' in
  \emph{CVPR}, 2017, p.~4.

\bibitem{CyclicGAN}
J.-Y. Zhu, T.~Park, P.~Isola, and A.~A. Efros, ``Unpaired image-to-image
  translation using cycle-consistent adversarial networks,'' in \emph{IEEE
  International Conference on Computer Vision}, 2017, pp. 2242--2251.

\bibitem{Photo-to-Caricature}
Z.~Zheng, C.~Wang, Z.~Yu, N.~Wang, H.~Zheng, and B.~Zheng, ``Unpaired
  photo-to-caricature translation on faces in the wild,''
  \emph{Neurocomputing}, vol. 355, pp. 71--81, 2019.

\bibitem{triplefaceimage}
L.~Ye, B.~Zhang, M.~Yang, and W.~Lian, ``Triple-translation gan with
  multi-layer sparse representation for face image synthesis,''
  \emph{Neurocomputing}, vol. 358, pp. 294--308, 2019.

\bibitem{multi-image-inpainting}
L.~Jiao, H.~Wu, H.~Wang, and R.~Bie, ``Multi-scale semantic image inpainting
  with residual learning and gan,'' \emph{Neurocomputing}, vol. 331, pp.
  199--212, 2019.

\bibitem{saliency}
Y.~Ji, H.~Zhang, and Q.~J. Wu, ``Saliency detection via conditional adversarial
  image-to-image network,'' \emph{Neurocomputing}, vol. 316, pp. 357--368,
  2018.

\bibitem{csgan}
K.~B. Kancharagunta and S.~R. Dubey, ``Csgan: Cyclic-synthesized generative
  adversarial networks for image-to-image transformation,'' \emph{arXiv
  preprint arXiv:1901.03554}, 2019.

\bibitem{cdgan}
K.~K. Babu and S.~R. Dubey, ``Cdgan: Cyclic discriminative generative
  adversarial networks for image-to-image transformation,'' \emph{arXiv
  preprint arXiv:2001.05489}, 2020.

\bibitem{karras2019style}
T.~Karras, S.~Laine, and T.~Aila, ``A style-based generator architecture for
  generative adversarial networks,'' in \emph{Proceedings of the IEEE
  Conference on Computer Vision and Pattern Recognition}, 2019, pp. 4401--4410.

\bibitem{mullick2019generative}
S.~S. Mullick, S.~Datta, and S.~Das, ``Generative adversarial minority
  oversampling,'' in \emph{Proceedings of the IEEE International Conference on
  Computer Vision}, 2019, pp. 1695--1704.

\bibitem{PerceptualGAN}
C.~{Wang}, C.~{Xu}, C.~{Wang}, and D.~{Tao}, ``Perceptual adversarial networks
  for image-to-image transformation,'' \emph{IEEE Transactions on Image
  Processing}, vol.~27, no.~8, pp. 4066--4079, Aug 2018.

\bibitem{dualGAN}
Z.~Yi, H.~Zhang, P.~Tan, and M.~Gong, ``Dualgan: Unsupervised dual learning for
  image-to-image translation,'' in \emph{IEEE International Conference on
  Computer Vision}, 2017, pp. 2868--2876.

\bibitem{ps2man}
L.~Wang, V.~Sindagi, and V.~Patel, ``High-quality facial photo-sketch synthesis
  using multi-adversarial networks,'' in \emph{IEEE International Conference on
  Automatic Face \& Gesture Recognition}, 2018, pp. 83--90.

\bibitem{vfsgan}
H.~Zhang, V.~M. Patel, B.~S. Riggan, and S.~Hu, ``Generative adversarial
  network-based synthesis of visible faces from polarimetrie thermal faces,''
  in \emph{2017 IEEE International Joint Conference on Biometrics
  (IJCB)}.\hskip 1em plus 0.5em minus 0.4em\relax IEEE, 2017, pp. 100--107.

\bibitem{tvgan}
T.~Zhang, A.~Wiliem, S.~Yang, and B.~Lovell, ``Tv-gan: Generative adversarial
  network based thermal to visible face recognition,'' in \emph{2018
  International Conference on Biometrics (ICB)}.\hskip 1em plus 0.5em minus
  0.4em\relax IEEE, 2018, pp. 174--181.

\bibitem{MTVFISGGAN}
C.~Chen and A.~Ross, ``Matching thermal to visible face images using a
  semantic-guided generative adversarial network,'' \emph{arXiv preprint
  arXiv:1903.00963}, 2019.

\bibitem{PAN}
C.~Wang, C.~Xu, C.~Wang, and D.~Tao, ``Perceptual adversarial networks for
  image-to-image transformation,'' \emph{IEEE Transactions on Image
  Processing}, vol.~27, no.~8, pp. 4066--4079, 2018.

\bibitem{LSGAN}
X.~Mao, Q.~Li, H.~Xie, R.~Y. Lau, Z.~Wang, and S.~P. Smolley, ``Least squares
  generative adversarial networks,'' in \emph{IEEE International Conference on
  Computer Vision}, 2017, pp. 2813--2821.

\bibitem{context_enc}
D.~Pathak, P.~Krahenbuhl, J.~Donahue, T.~Darrell, and A.~A. Efros, ``Context
  encoders: Feature learning by inpainting,'' in \emph{Proceedings of the IEEE
  conference on computer vision and pattern recognition}, 2016, pp. 2536--2544.

\bibitem{adam}
D.~P. Kingma and J.~Ba, ``Adam: A method for stochastic optimization,''
  \emph{International Conference on Learning Representations}, 2014.

\bibitem{TVFITGAN}
Z.~{Wang}, Z.~{Chen}, and F.~{Wu}, ``Thermal to visible facial image
  translation using generative adversarial networks,'' \emph{IEEE Signal
  Processing Letters}, vol.~25, no.~8, pp. 1161--1165, Aug 2018.

\bibitem{SSIM}
Z.~Wang, A.~C. Bovik, H.~R. Sheikh, and E.~P. Simoncelli, ``Image quality
  assessment: from error visibility to structural similarity,'' \emph{IEEE
  transactions on image processing}, vol.~13, no.~4, pp. 600--612, 2004.

\bibitem{LPIPS}
R.~Zhang, P.~Isola, A.~A. Efros, E.~Shechtman, and O.~Wang, ``The unreasonable
  effectiveness of deep features as a perceptual metric,'' in \emph{Proceedings
  of the IEEE Conference on Computer Vision and Pattern Recognition}, 2018, pp.
  586--595.

\bibitem{MSSSIM}
Z.~{Wang}, E.~P. {Simoncelli}, and A.~C. {Bovik}, ``Multiscale structural
  similarity for image quality assessment,'' in \emph{The Thrity-Seventh
  Asilomar Conference on Signals, Systems Computers, 2003}, vol.~2, Nov 2003,
  pp. 1398--1402 Vol.2.

\end{thebibliography}

\begin{IEEEbiography}[{\includegraphics[width=1in,height=1.25in,clip,keepaspectratio]{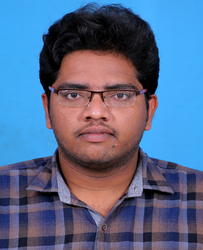}}]{Kancharagunta Kishan Babu}
is currently working toward the Ph.D. degree in Indian Institute of Information Technology, Sri City. His research interest includes Computer Vision, Deep Learning and Image Processing. He received the B.Tech. and M.Tech. degrees in Computer Science and Engineering from Bapatla Engineering College in 2012 and University College of Engineering Kakinada (JNTUK) in 2014, respectively.
\end{IEEEbiography}

\begin{IEEEbiography}[{\includegraphics[width=1in,height=1.25in,clip,keepaspectratio]{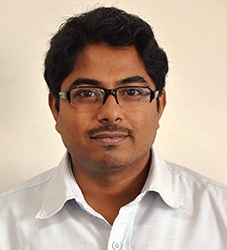}}]{Shiv Ram Dubey}
has been with the Indian Institute of Information Technology (IIIT), Sri City since June 2016, where he is currently the Assistant Professor of Computer Science and Engineering. He received the Ph.D. degree in Computer Vision and Image Processing from Indian Institute of Information Technology, Allahabad (IIIT Allahabad) in 2016. Before that, from August 2012-Feb 2013, he was a Project Officer in the Computer Science and Engineering Department at Indian Institute of Technology, Madras (IIT Madras).
He was a recipient of several awards including the Indo-Taiwan Joint Research Grant from DST/GITA, Govt. of India, Best PhD Award in PhD Symposium, IEEE-CICT2017 at IIITM Gwalior, Early Career Research Award from SERB, Govt. of India and NVIDIA GPU Grant Award Twice from NVIDIA. He received Outstanding Certificate of Reviewing Award from Information Fusion, Elsevier in 2018. He also received the Best Paper Award in IEEE UPCON 2015, a prestigious conference of IEEE UP Section.
His research interest includes Computer Vision, Deep Learning, Image Processing, Biometrics, Medical Imaging, Convolutional Neural Networks, Image Feature Description, Content Based Image Retrieval, Image-to-Image Transformation, Face Detection and Recognition, Facial Expression Recognition, Texture and Hyperspectral Image Analysis.
\end{IEEEbiography}

\end{document}